# Loud and Trendy: Crowdsourcing Impressions of Social Ambiance in Popular Indoor Urban Places

Darshan Santani
Idiap Research Institute
EPFL, Lausanne, Switzerland
dsantani@idiap.ch

Daniel Gatica-Perez
Idiap Research Institute
EPFL, Lausanne, Switzerland
gatica@idiap.ch

## ABSTRACT

New research cutting across architecture, urban studies, and psychology is contextualizing the understanding of urban spaces according to the perceptions of their inhabitants. One fundamental construct that relates place and experience is ambiance, which is defined as "the mood or feeling associated with a particular place". We posit that the systematic study of ambiance dimensions in cities is a new domain for which multimedia research can make pivotal contributions. We present a study to examine how images collected from social media can be used for the crowdsourced characterization of indoor ambiance impressions in popular urban places. We design a crowdsourcing framework to understand suitability of social images as data source to convey place ambiance, to examine what type of images are most suitable to describe ambiance, and to assess how people perceive places socially from the perspective of ambiance along 13 dimensions. Our study is based on 50,000 Foursquare images collected from 300 popular places across six cities worldwide. The results show that reliable estimates of ambiance can be obtained for several of the dimensions. Furthermore, we found that most aggregate impressions of ambiance are similar across popular places in all studied cities. We conclude by presenting a multidisciplinary research agenda for future research in this domain.

**Categories and Subject Descriptors:** H.4.m [Information Systems Applications]: Miscellaneous

## 1. INTRODUCTION

Cities are unique expressions of human activity and, at their core, are the intersection of physical spaces and the people who live in them. Cities are buildings and roads, but also the people who use them and create new knowledge through their continuous interaction and exchange of ideas [15]. Public places in cities have played a central role in facilitating a socio-cultural habitat for people to counterbalance the grind of daily life, an environment away from home and work [42, 6].

In this context, new research cutting across architecture, urban studies, and psychology is contextualizing the understanding of the urban space according to the perceptions of their inhabitants, which are rooted in both socio-economic factors and psychological constructs [14]. Studying cities from this multidisciplinary perspective is fundamental, as research is finding connections between psychological features of cities and key indicators like well-being and prosperity [29, 1]. Given the inevitable growth of urban life worldwide and the connections between place and well-being, an urgent goal is the development of scientific methodologies to provide "a better idea of how people perceive and experience places" [14].

One fundamental construct that relates place and experience is ambiance, defined as "the mood or feeling associated with a particular place" [26] or "the character and atmosphere of a place" [12]. As soon as we walk into a place, we can tell if it is made for us. We ubiquitously judge restaurants, cafes, or bars according to their ambiance – whether the venue is energetic, bohemian, loud, or trendy. In other words, we form place impressions combining perceptual cues that involve most senses (vision and hearing, but also smell, taste, and touch) as well as prior knowledge of both the physical space and its inhabitants [16]. As urban dwellers, we rely on ambiance to make decisions that have long-term impact, defining our favourite social hangouts and shaping our discoveries, including the kind of people we might end up meeting and interacting with.

The understanding of place ambiance in cities is a brave new topic for which multimedia research can play a pivotal role for several reasons:

1. Social media represents a crowdsourced mechanism to document urban places like restaurants, bars, clubs, and cafes. Due to the growth of sensor-rich mobile devices, online directory services like Yelp, Foursquare, and TripAdvisor are all popular today. These platforms provide users with functionalities to search for places in a given region and to leave feedback in the form of reviews and comments about their experience [3]. These services co-exist with integrated services like Google Places, Facebook Places, and Instagram. In all of them, users often take photos at venues and share them publicly. As a result, millions of images documenting places across the globe are available. The move by Google to acquire Zagat (one the most prestigious restaurant review services) and Hoppit service [19] are two examples of the increasing interest in industry to provide their users with detailed descriptions of places.

2. There is previous work in urban studies, architecture, psychology, and design that has studied connections between places and human perception, and that show that ambiance is a human way of relating to places [2, 17, 35]. This provides theoretical grounds for new computational research. However, most previous studies in these disciplines have faced the challenge represented by the difficulty to obtain place impressions. A standard approach involves physically visiting a place and gather impressions making silent observations about its atmosphere. Clearly, this approach is neither scalable nor captures contextual aspects of venues like time, by which a place might be ideal for a business lunch, but then turns into a trendy loud bar at night. In contrast, gathering place impressions based on images or videos shared on social media, where observers rate ambiance after viewing media items coming from a venue, has the advantages of being scalable, allowing the study of contextual factors like time, and spanning national boundaries to examine geographic and cultural differences in ambiance perception.

3. A holistic study of indoor place ambiance would involve the interpretation of multiple sources of information (images and videos, sound, text, and contextual metadata) in socially meaningful ways. Each of these modalities involve real challenges. As an example, visual cues used by venue owners to convey ambiance (and by patrons to perceive it) can be subtle, including colour schemes, lighting, spatial design, wall decorations and artwork, flooring and carpeting [2, 35, 18]. These visual categories pose many challenges for state-of-art computer vision algorithms. Similar points can be made for text or audio analysis.

4. Multimedia has over two decades of history of research on recognition of real-life visual scenes, moving from the original indoor/outdoor approaches in the late 1990s [36] to the recognition of functional place categories (like offices, corridors, etc) in the 2000s [33], and the recent use of Google Street View or Panoramio images to discover salient visual features in outdoor scenes in cities [13, 41]. Research on ubiquitous computing has also contributed to functional place categorization by focusing on other data inputs [24, 8]. Multimedia research has a younger history of understanding social and affective constructs from people [5] and media [10]. This said, little has been done so far to develop methodologies that allow to conceptualize and operationalize multimedia approaches to study social perception of urban places in general, and social ambiance of indoor places in particular.

We argue that a multimedia research agenda to study place ambiance is needed. Relying on a combination of social media data collection, image crowdsourcing, and data analysis, we contribute research resources and findings towards the understanding of how social ambiance can be systematically studied by the multimedia community. We address two research questions:

**RQ1:** What types of social media images best convey the ambiance of popular indoor places?

**RQ2:** Can the ambiance of an indoor place be reliably assessed by observers of social media images? If so, for what dimensions of ambiance?

The paper has five contributions:

1. Using Foursquare, we collect 50,000 images from 300 popular places from six metropolitan cities worldwide in three world regions – North America (New York City, Seattle, and Mexico City), Europe (Barcelona and Paris) and Asia (Singapore) (Section 3). In addition to geographic and cultural diversity, these cities are chosen because of their active user population on Foursquare. The focus of our collection is on *popular* places in Foursquare rather than on arbitrary places (see examples in Figure 1).

2. We design an image crowdsourcing experiment on Amazon's Mechanical Turk (MTurk) to assess the suitability of image categories that are most appropriate to describe the ambiance of a place (Section 4). Research in both psychology and computing has confirmed the feasibility of crowdsourcing as a way to conduct behavioral studies when appropriate incentives and mechanisms for quality control are established [22]. Using statistical tests, the results show that images with clear views of the environment are perceived by crowdworkers as being more informative of ambiance than other image categories, like food and drinks, or groups of people. Based on the crowdsourcing results, we build a refined image corpus suitable for indoor ambiance characterization.

3. A priori, the social ambiance of places is not known to zero-acquaintance visitors or observers. We design a second crowdsourcing experiment to assess how people perceive places from the perspective of ambiance (Section 5). We asked crowdworkers to rate indoor ambiance along 13 different physical and psychological dimensions where images served as stimuli to form place impressions. The ambiance categories include romantic, bohemian, formal and trendy, among others.

4. Based on the results obtained from the latter experiment, we find that reliable estimates of ambiance can be obtained using social media images, suggesting the presence of strong visual cues to form place impressions. Furthermore, while we identify a few statistically significant differences across cities along four ambiance dimensions, most aggregate impressions of ambiance are similar across popular places in all cities, which open relevant questions about the roles that geography and background knowledge of observers might be playing (Section 6).

5. We conclude by presenting a research agenda for future research in this domain along multiple axes of interest, some of which are multidisciplinary (Section 7). The resources we generated (data and annotations) are available to the research community. Overall, our work contributes to define multimedia approaches to understand the social perception of urban places in cities.

## 2. RELATED WORK

Given the multifaceted nature of our research questions, we review the related work along five axes: ubiquitous and multimedia computing, social media, hospitality research, social psychology, and urban computing.

The existing work on place characterization in ubiquitous computing, computer vision, and audio processing has examined several aspects including physical properties of places like their geographic location [21]; place composition, including the scene layout and the objects present in the scene [27]; place function, i.e., home, work, or leisure places; and place occupancy and noise levels [37]. This research has used both automatic [24, 8] and semi-automatic approaches [39] and a variety of data sources often studied in isolation, including images, sensor data like GPS/Wifi, and RF data. Works like [24, 8] have used audio or audio-visual data to characterized places through phone apps. The studied place categories (personal places in [24], home, work in [8]) differ significantly from ours. A recent work [37] investigates the recognition of physical ambiance categories (occupancy, human chatter, noise and music levels) using standard audio features collected in-situ by users. In contrast, our work examines social images as source of data, impressions of people who are not physically at the places, and a much larger number of social ambiance categories.

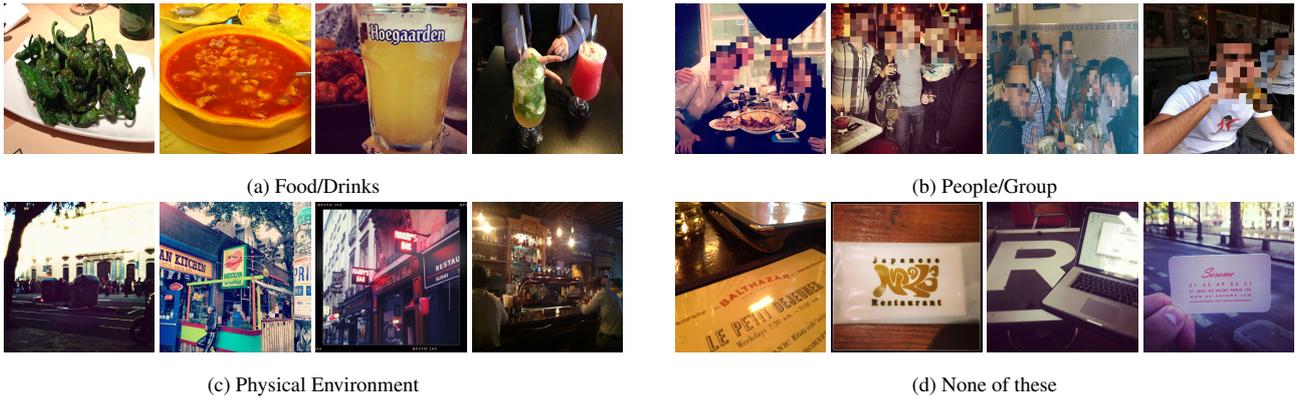

Figure 1: Sample images from *Random Image* corpus. Based on online annotations, a random set of four images which were classified as (a) Food/Drinks, (b) People/Group, (c) Physical Environment, and (d) None of these. For privacy reasons, images showing faces have been pixelated.

The emergence of social multimedia platforms, which allow users to take and share photos using mobile devices, have gained wide spread adoption. In the social multimedia literature, the work in [40] studies the problem of recommending locations based on mobility traces extracted from GPS and social links, without using image information. In contrast, the work in [7] uses geo-localized images, travel blog text data, and manual user profiles to suggest trips. Other works involving social images include [13] and [9]. These cases are focused on outdoor places and do not address the atmosphere dimensions we studied here. Due to the availability of large amount of images on Flickr or Instagram, researchers have analyzed these platforms (recent examples include [20, 4]). In [4], using a corpus of one million Instagram images, the authors studied the relationship between photos containing a face and its social engagement factors and found that photos with faces are more likely to receive likes and comments. As it relates to our work, an interesting result is that only 20% of images were found to contain faces, which suggests that many other image categories (related to food, places, etc.) exist (for an example, refer to the small-scale coding study in [20]).

In hospitality and retail studies, there has been significant interest to examine the effect of physical ambiance cues or "atmospherics" such as color, lighting, layout, and furnishing on customer perception and quality inferences [2, 35]. In a study conducted in a retail store, it was found that ambient (such as music, lighting, smell), design (such as color, ceilings, spatial layout) and social factors present in the store environment contribute towards higher merchandise and service quality [2]. In another study [18] the role of atmospherics across 10 full-service restaurants in Hong Kong was investigated. Using five dimensions of restaurant atmospherics (facility aesthetics, ambiance, spatial layout, employee factors, and view from the window) it was found that these dimensions have a significant influence on patrons' dining experience, and their willingness to pay more and recommend the restaurant to others. Similar results were obtained in another related work conducted across ethnic restaurants in the U.S. [23]. However, most of these studies are either done on controlled laboratory settings or based on questionnaires, which may have limitations with respect to ecological validity or recall biases.

Unlike the above research, we take a different direction examining the social perception of places (the ambiance impressions that people form about venues). Our proposed research is most closely related to work in social psychology [29, 16] which has investigated first impressions of places in connection to the personality of their inhabitants, mostly in controlled settings. A key first study [17] investigated the reliability (in terms of inter-rater agreement) of impressions of place ambiance and patron personality formed by (a) observers physically present at a number of indoor places, and (b) observers of Foursquare user profiles who visited those places (as opposed to views of places as we do). The results suggest that people do indeed form consistent impressions of ambiance and patrons traits. The study, however, only examined 49 places in one city (Austin, TX) and involved personally visiting every venue. In contrast, we study places in six cities.

An urban computing study [31] measured the perception of three variables (safety, class, and uniqueness) using 4,136 geo-tagged *outdoor* images in four cities, two each in the US and Austria. Images for the US cities were obtained via Google Street View, while manual collection was performed for the Austrian cities. The authors claimed differences in the range of perceptions of these dimensions between American cities and their Austrian counterpart. In a similar study on *outdoor* urban perception, judgments from over 3000 individuals were collected to study simple visual cues that could correlate outdoor places with three dimensions (beauty, quietness, and happiness) [28]. In both studies, dedicated websites were used to collect annotations, as opposed to common crowd-sourcing platforms like MTurk. Our research differs from these previous work on two specific grounds. First, we are interested in examining indoor places as opposed to outdoor spaces. These are clearly different categories from the urban design and urban studies perspectives. Second, we study 13 dimensions of social ambiance (including artsy, bohemian, loud, trendy, romantic, etc.), appropriate for the indoor setting, as opposed to dimensions studied in [31, 28], which reflect pedestrian or street-level characteristics.

## 3. SELECTION OF PLACES AND IMAGES

In this section we describe our methodology to select the list of popular places and their associated images across six cities.

### 3.1 Place Selection

We ground our analysis on data collected from Foursquare, a popular location-based social network. In Foursquare, users typically visit a place, announce their arrival (*check-in*) and share information about their visits to places with their friend circle. As per Foursquare rules, a place or a venue is a geographical location with fixed spatial coordinates, i.e., latitude and longitude. Throughout this paper, we will use place and venue interchangeably in the context of Foursquare.

| City | Ratings | Photos | Visitors |
|---|---|---|---|
| Barcelona | 8.66 (0.67) | 309.58 (383.53) | 1,874.34 (2,371.43) |
| NYC | 9.31 (0.41) | 463.62 (387.31) | 8,272.16 (6,208.76) |
| Paris | 8.55 (0.63) | 220.98 (254.16) | 1,685.76 (1,433.14) |
| Seattle | 8.95 (0.38) | 240.7 (147.94) | 3,533.54 (1,815.34) |
| Singapore | 8.29 (0.86) | 304.88 (206.58) | 3,457.64 (3,916.89) |
| Mexico City | 8.78 (0.49) | 361.34 (374.85) | 3,692.56 (3,578.84) |

Table 1: Summary statistics of Foursquare data. For each city, mean scores of attributes of popular places is shown, along with their standard deviations (shown in brackets.)

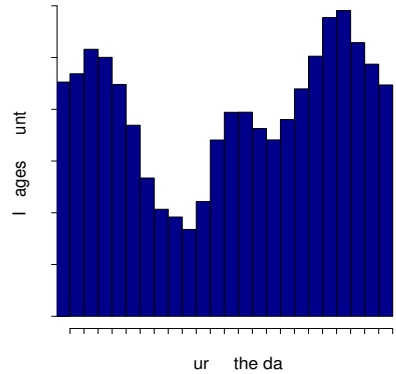

Figure 2: Histogram showing the frequency of images taken during different times of the day.

Each place on Foursquare maintains a profile page, which contains general information about the place (address, directions, phone number, etc.), in addition to Foursquare-specific data such as its popularity, total number of check-ins and past visitors, and a collection of images uploaded by users. Foursquare allows developers to obtain most of this information using its public API, which we used to gather all the relevant information for a given place [1].

For our analysis, we studied popular places on Foursquare for six cities around the world – Barcelona, Mexico City, New York City, Paris, Seattle, and Singapore. These cities were chosen for two reasons. First, they all are large cities in diverse world regions, and are known to have a vibrant urban life. Second, they all have an active user population on Foursquare. For each city, we chose 50 of the most popular places in each city which fall under the Foursquare-defined category of either being "Food" or "Nightlife Spots", which means cafes, restaurants, or bars. Table 1 lists the mean values of Foursquare data for all 50 places in each city. As stated in Section 1, we are focused on studying popular indoor places as opposed to arbitrary places (i.e., indoor or outdoor, and that might or might not be represented on Foursquare.)

Place selection was performed manually by the first author, taking into account place popularity, number of checkins, number of past visitors and number of available images. As image quality was an important criterion while selecting a place, we ignored all popular places which did not had any good-quality images such as dark images. In Table 1, using data obtained from Foursquare API, we notice that the user-generated mean rating of places selected for our study is above 8.2 (on a scale from 1 to 10) for all cities, confirming the popularity of places. Moreover, we also observe that these places are frequently visited by a large visitor population.

## 3.2 Image Selection

The second important consideration was the selection of images for each chosen place. We decided to select a small number of images per place to illustrate the place's atmosphere. This decision was motivated by the need to account for the variability in image quality, while at the same time providing general views of a place, without complicating the annotation process. Moreover, having more than one image for a place gives us the flexibility to show the place at different times of the day. Our hypothesis is that images of the physical environment will often be more representative of the place ambiance, so to test our hypothesis we built two image corpora, which are described in the next subsection.

[1] Foursquare has changed its mobile application and API significantly since our data collection.

### 3.2.1 Random Image Corpus

Given that each place in our database has on average more than 300 user-contributed images (see Table 1), it is challenging to select a small number of representative images that can accurately describe the ambiance of a place. One approach is to randomly select them given the collection of all images available for a place. We follow this approach to build a *random image* corpus.

Images for a place listed on Foursquare can be obtained via the API, but due to rate limits imposed by the API, we have access to at most 200 publicly visible images per place. We gathered a total of 50,023 images for all 300 places. This gives an average of 166 images per place, which is lower than the average estimated from the profile metadata ($\geq 300$), yet it remains a large number. In addition to gathering the images, the API also provides information on the image source (i.e., the application used to generate the photo), creation time, and other attributes such as image height and width. However, due to API restrictions we had access to metadata information for only 47,980 images.

Using the metadata information, we found that the median height and width of an image in our collected corpus is 720 pixels, with 55% of images taken via iPhone, 19% via an Android device, and 22% uploaded via Instagram. We also plot the distribution of image creation times in Figure 2. We identify three distinctive peaks – the first one occurs during the lunch hour (11am–1pm), the second peak around dinner time (6–8pm), and the last one occurs after midnight and early hours of the morning (nightlife). This result confirms our intuition that social media images provide a well-suited medium to capture places during different times of the day.

For the study described in this section, since we are interested in only a few images to represent each place, we randomly sampled three images per place from our corpus, to build a *random image* corpus of 900 images for all 300 places. Refer to Figure 1 for a sample of selected images from this corpus.

### 3.2.2 Physical Environment Image Corpus

Our second approach is to build an image corpus with clear views of the physical environment. We manually select a small number of images per place that satisfy this condition. Although this task can potentially be automated, we have chosen to manually control the quality of data for the crowdsourcing experiment. The selection was performed by the first author after browsing through all the user-contributed images. During the process, we opted for images with a view clearly showing the space from different angles (with or without the presence of visitors.) To the best of our ability, we avoided images where one can potentially identify faces, to protect the privacy of individuals. Moreover, we ensured that images that showed the venue name or any other information that explicitly revealed the identity of the place were discarded e.g., an image showing Starbucks or Hard Rock Cafe logos, to reduce any bias while characterizing the place ambiance. See Figure 3 for a sample of selected images from this corpus. Note that all these attributes cannot be controlled for while choosing images randomly.

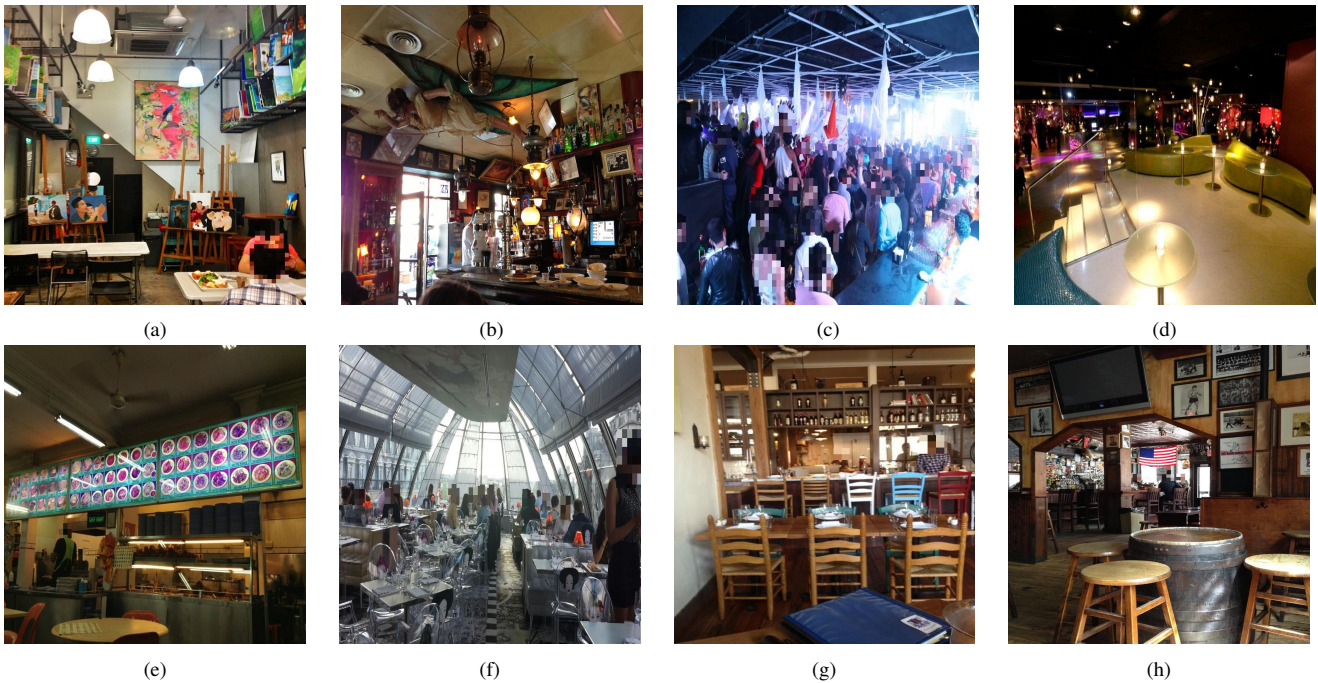

Figure 3: Sample images from *Physical Environment Image* corpus. Based on MTurk annotations, images which scored the **highest** on (a) Artsy, (b) Creepy, (c) Loud, and (d) Trendy; and **lowest** on (e) Artsy, (f) Creepy, (g) Loud, and (h) Trendy. For privacy reasons, images showing faces have been pixelated.

The *physical environment image* corpus also contains 900 images for all 300 places (three images per place). The manual selection was constrained by the quality of Foursquare images. At times, we encountered images which were not optimally bright or clear. However, this setting is realistic due to the absence of any beautified or vendor-provided images, which can potentially add biases to the impressions.

## 4. EXPERIMENT 1: SUITABILITY OF IMAGES TO CONVEY AMBIANCE

In this section we address RQ1, i.e., we use both image corpora to judge which approach results in better "ambiance quality", that is, images which are more adequate to convey ambiance according to crowd judgements. On one hand, random selection of images is a realistic "in the wild" setting that provides an automated way to collect images. However, it will represent a valid approach only if it results in a collection of images which provide sufficient visual cues to characterize place ambiance. On the other hand, the manual selection of physical environment images is a controlled setting that satisfies the criteria described in the previous paragraphs.

### 4.1 Crowdsourcing Image Impressions

Our hypothesis is that many images in the *random image* corpus might not be perceived by crowdworkers as ideal to characterize a place's ambiance, as they do not contain visual cues to gauge a place's physical environment. In our exploratory inspection, most of these random images contain food items or show groups of people. To answer RQ1, we conducted a crowdsourcing study to gather the perceived ability of both image corpora – *random* and *physical environment*, to describe a place's ambiance and physical environment. For crowdsourcing, we used MTurk and chose US-based workers with at least 95% approval rate for historical HITs (Human Intelligence Tasks). In addition, to increase the potential reliability of MTurk annotations, we only chose "Master" annotators, which typically involves a worker pool with an excellent track record of completing tasks with precision.

For each HIT annotation task, we picked a set of five images per place, consisting of two from the *physical environment image* corpus, and three from the *random image* corpus. We ensured that images from the two sets do not overlap. In each HIT, workers were asked to view these five images and then answer three questions. In the first question, the workers were asked to rank the images based on how informative they were of the ambiance of the place. In the second one, workers were asked to rank the same set of images based on their degree of information about the physical environment of the place. The third question asked workers to categorize the images in four classes: a) Food/Drinks, b) People/Group, c) Physical Environment, and d) None of these. For these questions, no explicit definitions of ambiance, physical environment, food/drinks or people were provided, as we wanted the workers to rely on their internal representation of these concepts.

In the two ranking questions, images cannot be given the same rank, each image needed to have a different rank. For the image categorization task, the workers were asked to choose exactly one category for each image. If the images had the same rank or fell into one or more categories, we asked the annotators to pick the rank or category that for them was the best choice. We collected 10 annotations for each HIT across all 300 places, for a total of 3,000 responses for every question.

We also gathered crowdworkers' demographics via an email-based survey. We asked workers about their age group, gender,

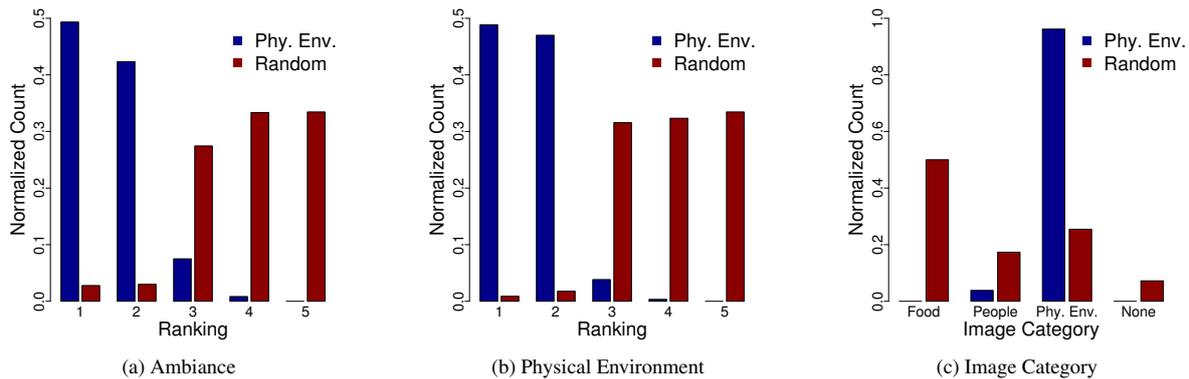

Figure 4: Results for *Majority Vote* aggregation technique. Plot showing the histograms for a) Ambiance, b) Physical Environment, and c) Image Category, for both *Physical Environment Image* and *Random Image* corpus.

| Method | Physical Environment | | | | Random | | | |
|---|---|---|---|---|---|---|---|---|
| | Top 2 | | Bottom 3 | | Top 2 | | Bottom 3 | |
| | Ambiance | Phy. Env. | Ambiance | Phy. Env. | Ambiance | Phy. Env. | Ambiance | Phy. Env. |
| Majority Vote | 91.7% | 95.8% | 8.3% | 4.2% | 5.8% | 2.7% | 94.2% | 97.3% |
| Median | 89.7% | 94.7% | 10.3% | 5.3% | 4.1% | 2.0% | 95.9% | 98.0% |

Table 2: Table showing the summary statistics for each aggregation method. For each method, we show the percentage of images from both image corpora which are either in Top 2 ranks (rank 1 or 2), or ranked in Bottom 3 (ranks 3,4,5).

education level, current place of residence (categorized as either rural, suburbs, small-sized town, mid-sized town or city), and any experience of living in a big city. We also inquired them about their typical frequency to go out for food or drinks (almost every day, 2-3 times per week, once a week, 1-2 times a month, or less than once a month). These questions were designed to understand the ability of workers to rate images for ambiance and physical environment based on previous experiences.

## 4.2 Results

In this section, we present the results of our first crowdsourcing experiment.

### 4.2.1 Worker Participation and Demographics

For a total number of 3,000 HITs available for this experiment, we observe that a typical worker completed an average of 39 HITs. While 50% of the workers submitted less than 9 HITs, the worker with the highest number of HITs completed 295 assignments. We observe a long-tailed distribution in HIT completion times (mean: 114 secs, median: 88 secs, max: 593 secs). It is worth noting that we allocated a maximum of 10 minutes per HIT.

We had a pool of 101 workers who responded to our HITs. Of all HIT respondents, 32% replied to our demographics survey. We notice a balanced gender ratio (50% of workers being female), which corroborates earlier findings in the literature [30]. While only 34% of our worker pool currently lives in a big city, 75% of them have already experienced city living in the past. Furthermore, 56% of them go out for food or drinks at least once a week. These findings provide evidence that the majority of respondents are likely capable to assess ambiance in urban environments. We also notice that the worker population is relatively not so young with the most popular category (53%) being the age group of 35-50 years old. Note that the worker demographics reported here encompasses the worker population in both crowdsourcing experiments of this paper.

### 4.2.2 Analysis of Annotations

Now we turn our focus towards assessing the suitability of each image corpus to convey ambiance. As mentioned earlier, for each HIT we collected 10 impressions per place. Thus, it becomes important to consider the role of different aggregation methods in analyzing the results. Aggregation is used to create a composite score per place given the 10 responses for each question. In other words, for every question, aggregation is performed at the place-level. We use two different aggregation techniques. The first one is the *majority vote*, where we compute the majority score given the 10 impressions for each place. We then summarize the results based on 300 majority impressions. For the *median* method, we compute the median as the composite score across 10 impressions for each place.

Table 2 lists the summary statistics for the two aggregation techniques. For each aggregation technique and each corpus, we report the percentage of images which are in Top 2 ranks (ranks 1,2) and Bottom 3 ranks (ranks 3,4,5). We list these statistics for both the ambiance and physical environment questions. For the *majority vote* technique, manually selected images (*physical environment image* corpus) are in Top 2 ranks 91.7% for ambiance and 95.8% for physical environment, while random images are in Bottom 3 ranks for 94.2% and 97.3%, respectively. Note that a random ranking method would assign the manually selected image in the Top 2 rank with a probability of $1/10$ $(1/\binom{5}{2})$. We also plot the histogram of rankings for image sets from both corpora in Figures 4a and 4b. These results show that manually selected images are associated with higher ranks, while the random set of images are associated with lower ranks for both ambiance and physical environment, irrespective of the aggregation technique.

In addition to asking annotators to rank images, we also asked them to classify images into one of the four categories (food/drinks, people/group, physical environment, and none of these), as described in Section 4.1. In Figure 4c, we plot the assigned category for the *majority vote* technique. We observe that images from the *physical environment image* corpus are labeled as describing the physical environment in 96.2% of the cases. In contrast, im-ages from the *random image* corpus are representative of either food/drinks, or people, or other in 74.6% of cases combined, and showing food items or people in 67% of cases.

| Label | Barcelona | New York City | Paris | Seattle | Singapore | Mexico City | Combined | Graham [17] |
|---|---|---|---|---|---|---|---|---|
| Artsy | 0.81 | 0.66 | 0.69 | 0.72 | 0.80 | 0.76 | 0.76 | 0.63 |
| Bohemian | 0.63 | 0.58 | 0.54 | 0.54 | 0.66 | 0.66 | 0.62 | 0.67 |
| Conservative | 0.67 | 0.77 | 0.78 | 0.67 | 0.70 | 0.85 | 0.76 | 0.77 |
| Creepy | 0.54 | 0.62 | 0.60 | 0.57 | **0.32** | 0.62 | 0.59 | 0.81 |
| Dingy | 0.74 | 0.81 | 0.59 | 0.69 | 0.67 | 0.81 | 0.74 | 0.74 |
| Formal | 0.76 | 0.93 | 0.93 | 0.89 | 0.89 | 0.90 | 0.91 | 0.82 |
| Sophisticated | 0.68 | 0.91 | 0.90 | 0.85 | 0.80 | 0.87 | 0.86 | 0.70 |
| Loud | 0.80 | 0.81 | 0.76 | 0.74 | 0.82 | 0.82 | 0.80 | 0.74 |
| Old-fashioned | 0.82 | 0.46 | 0.78 | 0.45 | 0.72 | 0.67 | 0.72 | 0.67 |
| Off the beaten path | 0.58 | 0.62 | 0.39 | 0.54 | 0.61 | 0.59 | 0.58 | 0.73 |
| Romantic | 0.38 | 0.84 | 0.86 | 0.80 | 0.83 | 0.86 | 0.82 | 0.63 |
| Trendy | 0.69 | 0.71 | 0.50 | 0.43 | 0.68 | 0.85 | 0.69 | 0.58 |
| Up-scale | 0.69 | 0.91 | 0.90 | 0.85 | 0.81 | 0.85 | 0.86 | 0.76 |

Table 3: $ICC(1, k)$ scores of 13 ambiance dimensions for each city. $ICC(1, k)$ scores obtained in [17] are also shown in the last column for comparison. Cells marked in **bold** are **not** statistically significant at $p < 0.01$.

*Statistical Comparison*: We perform a statistical comparison of rankings between both image corpora for ambiance and physical environment dimensions. We performed the Wilcoxon signed-rank statistical test, which is a non-parametric test to compare the mean ranks of two populations [38]. At the 99% confidence level, we obtained a *p*-value $< 2.2 \times 10^{-16}$ for both dimensions, validating the hypothesis that manually selected images are perceived by crowd-workers as better describing ambiance and physical environment.

In summary, these results provide an answer to RQ1, validating that images with clear views of the environment from Foursquare places are perceived as more suitable to characterize indoor ambiance than other image categories, as they contain visual cues to gauge a place's ambiance and physical environment. Please note that in this section we report the summary statistics across all cities combined. Individual trends for each city are similar to the overall trends and are omitted due to space restrictions.

## 5. EXPERIMENT 2: DIMENSIONS OF PLACE AMBIANCE

A priori, the ambiance of places is not known to zero-acquaintance observers. In this section, we address our second question (RQ2) i.e., whether reliable estimates of ambiance can be obtained using Foursquare images. Based on the *physical environment image* corpus of 900 images across 300 places, we design a second crowdsourcing experiment and asked crowd workers to rate indoor ambiance along 13 different physical and psychological dimensions where images served as stimuli to form place impressions.

### 5.1 Selection of Ambiance Categories

In order to select dimensions to characterize place ambiance, we base our methodology on prior work [17]. The authors proposed a rating instrument consisting of 41 dimensions for ambiance characterization. In our work, we chose 13 dimensions for which the corresponding intraclass correlations were amongst the highest as reported in [17]. Note that many dimensions in [17] did not reach sufficient inter-annotator agreement. We used a five-point Likert scale ranging from *strongly disagree* (1) to *strongly agree* (5) to judge the ambiance labels, while [17] used a 3-point categorical scale (yes, maybe, no). We will use the terms dimensions and labels interchangeably in the context of ambiance categories. The list of selected labels is shown in alphabetical order in Table 3.

### 5.2 Crowdsourcing Ambiance Impressions

To answer RQ2, crowdsourcing was employed to gather ambiance impressions. We used MTurk with the same worker qualification requirements as the first study. In each HIT, the workers were asked to view three images corresponding to a place, and then rate their personal impressions of the place ambiance based on what they saw. As part of the annotation interface, we ensured that workers viewed images in high resolution (and not just the image thumbnails). People were given a previous definition of each ambiance category. Moreover, workers were not informed about the city under study to reduce potential bias and stereotyping associated to the city identity. We collected 10 annotations for each dimension across all 300 places, for a total of 3,000 responses.

### 5.3 Results

For the 3,000 available HITs in this experiment, a typical worker completed an average of 56 HITs, with one worker completing 270 HIT assignments. When compared to the first experiment, similar results were obtained for HIT completion times (mean: 97 secs, median: 68 secs, max: 596 secs).

We turn our focus towards assessing the reliability of the annotations. We measure the inter-annotator consensus by computing intraclass correlation (ICC) among ratings given by the worker pool. Our annotation procedure requires every place to be judged by $k$ annotators randomly selected from a larger population of $K$ workers ($k = 10$, while $K$ is unknown as we have no means to estimate the MTurk worker population). Consequently, $ICC(1, 1)$ and $ICC(1, k)$ values, which respectively stand for single and average ICC measures [32], are computed for each ambiance dimension across all cities.

Table 3 reports the $ICC(1, k)$ values for all cities (due to space constraints, we omit $ICC(1, 1)$ values.) In addition to listing the individual scores for each city and label, we also report the combined $ICC(1, k)$ scores for each label for the whole dataset, where we have combined all places across cities. We observe acceptable inter-rater reliability for many labels, with all the scores being statistically significant (*p*-value $< 0.01$), with the exception of *creepy* label in Singapore. Furthermore, we notice that the inter-rater reliability for labels *formal*, *sophisticated*, *romantic*, and *up-scale* is typically high (above 0.8) for most of the cities. Using correlation analysis between labels (which is presented in Section 6.2), we find that these four labels are collinear, with pairwise correlations exceeding 0.8. It is interesting to note that label *loud* achieved high agreement from images not showing any sound (0.8 combined score). On the other hand, labels *creepy* and *off the beaten path* are the labels with the lowest ICC (below 0.6 for the combined score.)

| Label | Barcelona | Mexico City | NYC | Paris | Seattle | Singapore | Combined |
|---|---|---|---|---|---|---|---|
| Artsy | 2.54 (0.78) | 2.20 (0.69) | 2.14 (0.56) | 2.36 (0.69) | 2.05 (0.59) | 2.46 (0.72) | 2.29 (0.69) |
| Bohemian | 2.34 (0.60) | 1.94 (0.58) | 2.07 (0.49) | 2.09 (0.55) | 1.99 (0.44) | 2.04 (0.57) | 2.08 (0.55) |
| Conservative | 2.04 (0.58) | 2.36 (0.81) | 2.33 (0.70) | 2.17 (0.71) | 2.37 (0.57) | 2.28 (0.59) | 2.26 (0.67) |
| Creepy | 1.33 (0.31) | 1.37 (0.38) | 1.21 (0.27) | 1.20 (0.27) | 1.21 (0.27) | 1.18 (0.19) | 1.25 (0.29) |
| Dingy | 1.68 (0.52) | 1.61 (0.60) | 1.60 (0.58) | 1.49 (0.40) | 1.57 (0.46) | 1.49 (0.42) | 1.57 (0.50) |
| Formal | 1.60 (0.50) | 2.13 (0.91) | 2.14 (0.97) | 2.01 (0.96) | 1.95 (0.75) | 1.62 (0.68) | 1.91 (0.84) |
| Sophisticated | 2.09 (0.56) | 2.41 (0.90) | 2.42 (0.96) | 2.37 (0.93) | 2.20 (0.74) | 2.15 (0.70) | 2.27 (0.82) |
| Loud | 2.30 (0.67) | 2.45 (0.78) | 2.51 (0.72) | 2.09 (0.68) | 2.33 (0.62) | 2.49 (0.83) | 2.36 (0.73) |
| Old-fashioned | 2.20 (0.77) | 2.30 (0.62) | 2.33 (0.46) | 1.90 (0.67) | 2.44 (0.41) | 2.16 (0.56) | 2.22 (0.61) |
| Off the beaten path | 2.27 (0.51) | 1.88 (0.53) | 2.06 (0.52) | 1.89 (0.46) | 1.99 (0.45) | 1.96 (0.48) | 2.01 (0.50) |
| Romantic | 1.77 (0.37) | 2.09 (0.81) | 1.95 (0.72) | 1.92 (0.78) | 1.86 (0.62) | 1.80 (0.65) | 1.90 (0.68) |
| Trendy | 2.34 (0.65) | 2.55 (0.89) | 2.55 (0.66) | 2.49 (0.55) | 2.45 (0.47) | 2.54 (0.64) | 2.49 (0.65) |
| Up-scale | 1.93 (0.56) | 2.36 (0.85) | 2.39 (0.93) | 2.36 (0.91) | 2.13 (0.70) | 2.01 (0.69) | 2.20 (0.80) |

Table 4: Means and standard deviations (in brackets) of annotation scores for each city and label.

Importantly, these reliability scores are comparable to the ones obtained by Graham et al. [17], who conducted a similar study, but where the raters physically visited every venue (see section 2 and last column of Table 3), while in our case online images act as a stimuli. To summarize, these results provide an answer to RQ2 as they suggest that consistent impressions of place ambiance can be formed based upon images contributed in social media, which further suggests that there might be strong visual cues present for annotators to form accurate place impressions. The investigation of what specific cues contribute to impression formation will be the subject of future work.

## 6. AMBIANCE ACROSS CITIES

In this section we present descriptive statistics and study differences across cities for each ambiance label.

### 6.1 Descriptive Statistics

Table 4 lists the descriptive statistics (mean score and standard deviation) for each city and label. The mean scores are derived as follows: first, for every place we compute the mean score for each ambiance label, using the 10 annotations per label for each place; we then compute the mean scores and standard deviations for each city and label using the 50 places for each city. At the level of individual annotations, minimum and maximum values are 1 and 5 respectively for all each city and label, showing that the full scale is used by the crowd-workers. Note that the mean value obtained for all labels and all cities is below 3, which indicates a trend towards disagreement with the corresponding label. On the other hand, each city has venues that score high for each dimension.

In all cities, except Barcelona, the mean score for *trendy* is the highest amongst all labels; Barcelona places score the maximum on being *artsy*. *Creepy* scores the lowest (along with the lowest variance) for all cities, which is not surprising given that all places are popular places in their respective cities. From Table 4, we do not observe much variation in the mean values across cities, but a few differences stand out. For instance, the mean differences of the *formal* attribute between NYC and Barcelona, and the *old fashioned* attribute between Paris and Seattle exceed 0.5, potentially suggesting differences in place perceptions. We explore this further in Section 6.3.

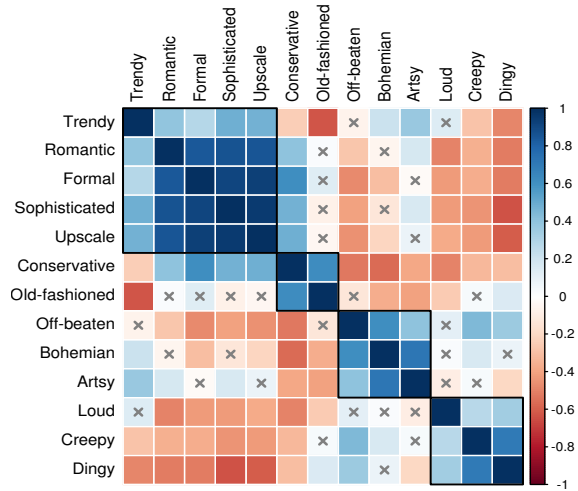

Figure 5: Plot showing the correlation matrix between ambiance dimensions. Black rectangular borders indicate the four distinct clusters found in the correlation matrix. Cells marked **X** are **not** statistically significant at $p < 0.01$.

### 6.2 Correlation Analysis

To look for linear relationship between ambiance labels, we perform correlation analysis using the mean annotation scores for all ambiance labels. Figure 5 visualizes the correlation matrix across all ambiance dimensions. We have used hierarchical clustering to re-order the correlation matrix in order to reveal its underlying structure. We color code the matrix instead of providing numerical scores to facilitate the discussion. We observe four distinct clusters. Starting from top left in the first cluster, labels *formal*, *sophisticated*, *romantic* and *up-scale* are highly collinear with pairwise correlations exceeding 0.8. The second cluster consists of places which are either *conservative* or *old-fashioned*, and the third cluster consists of *off-beaten*, *bohemian* or *artsy* places. The fourth cluster (bottom-right) lies on the opposite spectrum with respect to cluster one, and consists of *loud*, *dingy* and *creepy* places. Each of these four clusters clearly correspond to different ambiances. Furthermore, we can also observe significant negative correlations between dimensions in cluster one and cluster four and between clusters two and three.

| Label | City Pair | Mean Difference | $p$−value $\times 10^{-3}$ |
|---|---|---|---|
| Old Fashioned | SEA–PAR | $+0.544$ | $9.9 \times 10^{-2}$ ($7.4 \times 10^{-3}$) |
| Artsy | SEA–BCN | $-0.492$ | 4.26 (6.18) |
| Old Fashioned | PAR–NYC | $-0.434$ | 4.09 (0.67) |
| Bohemian | MEX–BCN | $-0.398$ | 3.70 (**39.68**) |
| Off the beaten path | MEX–BCN | $-0.386$ | 1.43 (0.051) |
| Off the beaten path | PAR–BCN | $-0.376$ | 2.11 (0.67) |

Table 5: Tukey's HSD and KS test statistics. $p$−values obtained from KS test are shown in brackets in the last column. Values marked in **bold** are **not** statistically significant at $p < 0.01$

## 6.3 Statistical Comparison

To better understand whether mean differences across cities for some of these ambiance labels are statistically significant, we perform the Tukey's honest significant difference (HSD) test. Tukey's HSD test is a statistical procedure for groups which compares all possible pairs of mean values for each group, the null hypothesis being that the mean values being compared are drawn from the same population [34]. We perform the HSD test to compute pairwise comparisons of mean values between cities for each ambiance label, which result in a total of 195 comparisons (15 city-wise pairs across 13 dimensions). Table 5 lists only the significant results of the Tukey's HSD test, where the differences in the observed means are statistically significant at $p$-value < 0.01. Based on these statistics and commenting only on results where differences is larger than 0.4, we observe that:

1. Popular places in Seattle are perceived as less *artsy* compared to places in Barcelona;

2. Popular places in Paris are perceived as less *old fashioned* compared to New York City and Seattle.

To validate the statistical significance of the Tukey's HSD test, we perform a series of pairwise Kolmogorov-Smirnov test (KS test) across all cities and labels. The KS test is a non-parametric test to compare the empirical distributions of two samples, with the null hypothesis being that the two samples are drawn from the same distribution [25]. We perform the KS test to compare the cumulative distribution functions of each city-pair across each dimension (195 comparisons). We report the $p$−values for the KS test in Table 5 for a statistical level $\alpha = 0.01$. Results from the KS test confirms most of the results from the Tukey's HSD test.

To conclude this subsection, our study shows that most of the differences across cities for each of the ambiance dimensions are not statistically significant. This result is interesting in itself as it might suggest that popular places in social media in cosmopolitan cities have many points in common. To our knowledge, this is a result that has not been reported before in social media research, but that could have some support from literature that discusses the "uniformization of taste" in globalized cities [11] and social media content. This said, any possible interpretation would have to be further validated with more data and a combination of further data analysis and ethnography. In addition, these results highlight the need to study other venues, including not so popular places on Foursquare and places not represented on Foursquare because of well-known socio-economic biases in social media.

## 7. AN AGENDA FOR FUTURE RESEARCH

We presented an emerging subject in social media, namely the study of psychological dimensions of the urban space from social media content under the construct of ambiance. Using over 50,000 Foursquare images collected from 300 popular indoor places across six cities, we first assessed the suitability of social images as data source to convey place ambiance, and found through a crowdsourcing experiment that images with clear views of the environment were significantly perceived by people as being more informative of ambiance than other image categories. Second, we demonstrated through another crowdsourcing experiment involving 13 ambiance dimensions that reliable estimates of ambiance for several dimensions can be obtained using Foursquare images, suggesting the presence of strong visual cues to form place impressions. Furthermore, we found that most aggregated impressions of popular places are similar across cities, with a few statistically significant differences across four ambiance dimensions (*bohemian*, *artsy*, *old fashioned*, and *off the beaten path*).

This topic extends multimedia research through the connection of "traditional" multimedia research with emerging themes like multimedia crowdsourcing and psychology of both social media and cities. We close the paper by defining a research agenda, which would include (and could extend beyond) the following issues:

1. The understanding of what specific cues in an indoor venue people use to judge a place and form ambiance impressions when looking at social media content – color, lighting scheme, spatial layout, interior design, or customers – is a fundamental open issue. This topic requires the integration of psychology research (e.g. models of cue utilization and validity) and of existing knowledge in hospitality, design, and marketing (known as atmospherics in those domains). Multimedia research could further use computer vision or audio processing algorithms to extract and reason about some of these cues.

2. The automatic recognition of ambiance is another key open issue. While there is a variety of potentially informative cues (visual cues from images or video, acoustic cues that could be collected on-site, text from comments, tips, and metadata), the specific connections of these features with ambiance still need to be established. We hypothesize that some of the studied dimensions have the potential to be automatically recognized, and would like to explicitly propose a challenge to the multimedia community on this topic.

3. The long tail (of non-popular places) is a third issue. The intrinsic biases of social media result in few places being richly represented (as the ones studied here), but the majority of urban indoor venues that are poorly represented (if at all) could be also studied under the ambiance lens. Could significant differences in impressions between popular and non-popular venues be quantified? Could one devise transfer learning approaches that could be used to learn from popular places and adapted for non-popular ones? These are just a couple of research questions that ought to be investigated.

4. The possibility of global studies of place ambiance is a fourth key dimension. Our results showed few significant differences among the impressions of observers across the six cities we studied. There is a need to expand and explain this concrete result, but more importantly there is open space to think about bigger questions that relate ambiance, culture, and economics at the collective level. Are the tastes of "global" social media users being uniformized, and if so is this being reflected in the media they collect and the judgements they make? Or are the socio-economic biases in social media hiding the diversity of venues that one could expect to exist across cities and cultures? Research in this domain could integrate multimedia analysis with cross-cultural studies.

5. Finally, the applications derived from the above research are manifold. On the user side, this could include hyper-local, ambiance-driven place search and discovery in online platforms, where users could search for places by its ambiance e.g., a *formal* place for a family dinner, or a *romantic* place for a date. This could complement existing sources of information like place reviews. As a second application domain, on the side of venue owners, this could include data-driven tools to deepen the understanding of the impressions that their venues evoke in potential and real customers, and recommendations about improving the appearance and style of their venues.

To conclude, as a concrete step towards advancing the above agenda, we are making the dataset (data and annotations) publicly available for research here: `https://www.idiap.ch/dataset/place-ambiance`

## 8. ACKNOWLEDGEMENTS

This work has been supported by the SNSF Youth@Night project. We thank anonymous reviewers for their valuable feedback and MTurk workers for their participation in our study.